\documentclass{ws-procs9x6}
\usepackage{overpic}
\usepackage{epsfig}

\newcommand{\CK}{\v Cerenkov}

\begin{document}

\title{Velocity and charge reconstruction with the AMS/RICH detector}
   
\author{\underline{Lu\'isa Arruda}, F.Bar\~ao, J.Borges, F.Carmo,\\
  P.Gon\c{c}alves, R.Pereira M.Pimenta}
\address{LIP/IST \\
         Av. Elias Garcia, 14, 1$^o$ andar\\
         1000-149 Lisboa, Portugal \\
         e-mail: luisa@lip.pt}

\maketitle

\abstracts{                                                    
The Alpha Magnetic Spectrometer (AMS), to be installed on the 
International Space Station (ISS) in 2008, will be equipped with a 
proximity focusing Ring Imaging \CK\ detector (RICH).
This detector will be equipped with a dual radiator (aerogel+NaF), 
a lateral conical mirror and a detection plane made of 680 photomultipliers
and light-guides, enabling measurements of particle electric charge and
velocity. A likelihood method for the \CK\ angle
reconstruction was applied leading to a velocity determination for protons
with a resolution around 0.1\%. The electric charge reconstruction is based on the counting
of the number of photoelectrons and on an overall efficiency estimation on an
event-by-event basis. Results from the application of both methods are
presented.} 
\vspace{-1.0cm}
\section{The AMS02 detector}
AMS~\cite{bib:ams} (Alpha Magnetic Spectrometer) is a precision spectrometer 
designed to search for cosmic antimatter, dark matter and to study the relative abundance of elements and isotopic composition of the primary cosmic rays. 
It will be installed in the International Space Station (ISS), in 2008, where
it will operate, at least, for a period of three years.

\begin{figure}[htb]
\begin{center}
\vspace{-1.cm}
\epsfig{file=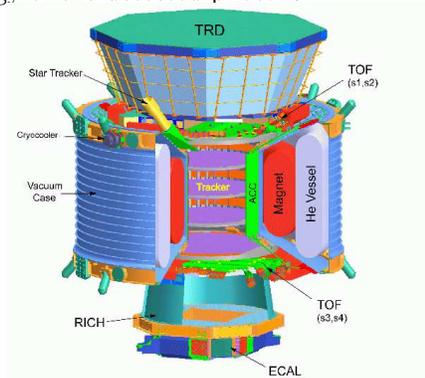,width=0.5\linewidth,clip=,bbllx=0,bblly=0,bburx=612,bbury=550}  
\vspace{-0.5cm} 
\caption{A whole view of the AMS Spectrometer.}
\end{center}  
\label{fig:ams}
\end{figure}

The spectrometer  will be capable of measuring the rigidity ($R\equiv pc/ |Z| e$), the charge ($Z$),
the velocity ($\beta$) and the energy ($E$) of cosmic rays within a 
geometrical acceptance of $ \sim$0.5\,m$^2$.sr.
Fig. \ref{fig:ams} shows a schematic view of the AMS spectrometer.
On top, a Transition Radiation Detector (TRD) 
will discriminate between leptons and hadrons.
It will be followed by the first of the four Time-of-Flight (TOF) system 
scintillator planes. The TOF will provide a fast trigger, 
charge and velocity measurements for charged particles, as well as information on their direction of incidence.
The tracking system will be surrounded by Veto Counters and
embedded in a magnetic field of about 0.9\,Tesla produced by a 
superconducting magnet.
It will consist on a Silicon Tracker, 
constituted of 8 double sided silicon 
planes, providing both charge and rigidity measurements with an accuracy
better than 2\% up to 20\,GV. The maximum detectable rigidity is around 1\,TV. 
The Ring Imaging \CK\ Detector (RICH), described in the next section,
will be located right after the last TOF plane and before the 
Electromagnetic Calorimeter (ECAL) which will enable $e/p$ separation 
and will measure the energy of the detected photons.\\

\begin{figure}[htb]
\begin{center}
\vspace{-.7cm}
\begin{tabular}{cc}  
\hspace{-.68cm}
\includegraphics[scale=0.28,angle=0,clip=,bb=14 14 599 396]{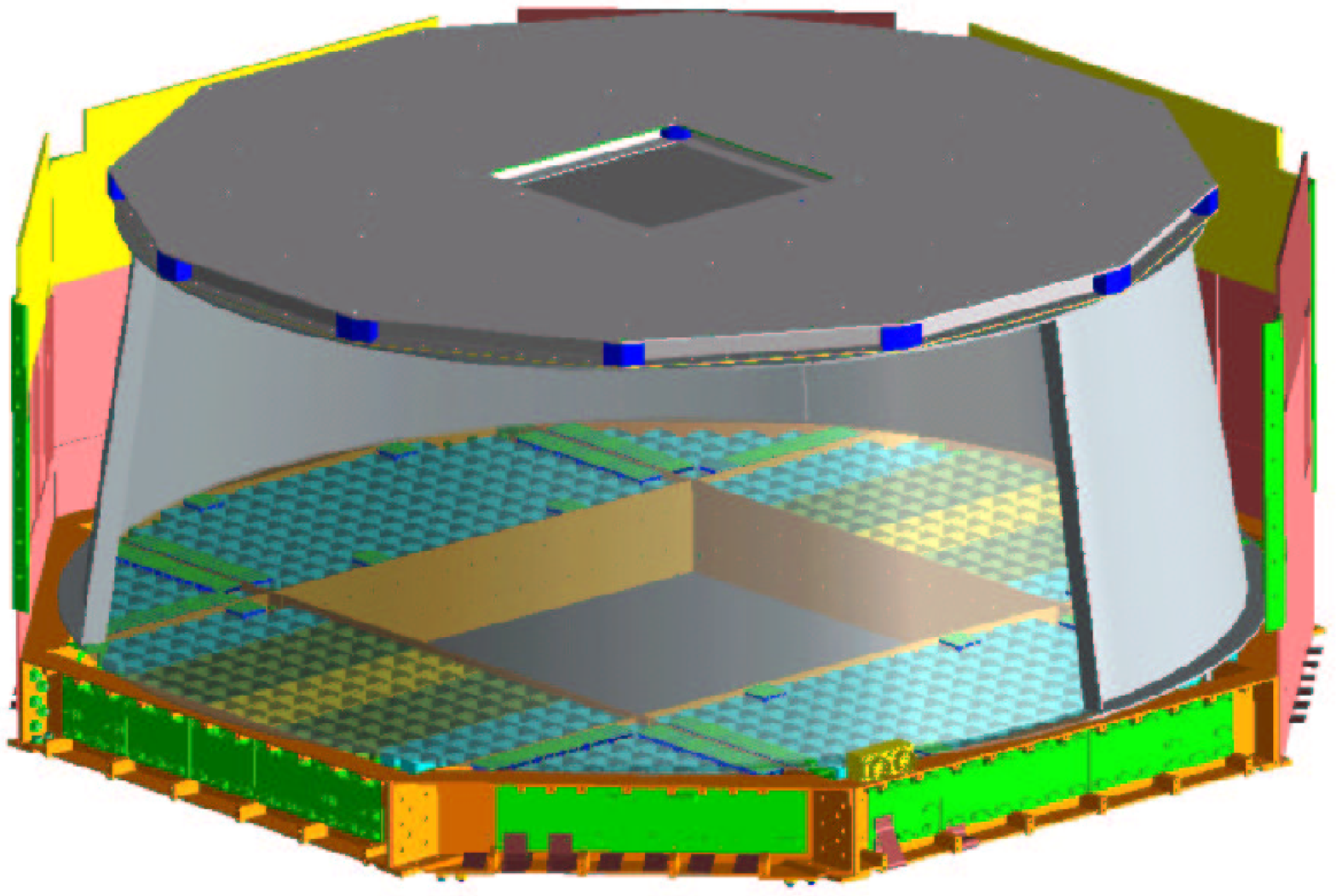}
&
\hspace{-.3cm}
\includegraphics[scale=0.27,angle=0,clip=,bb=14 14 496 463]{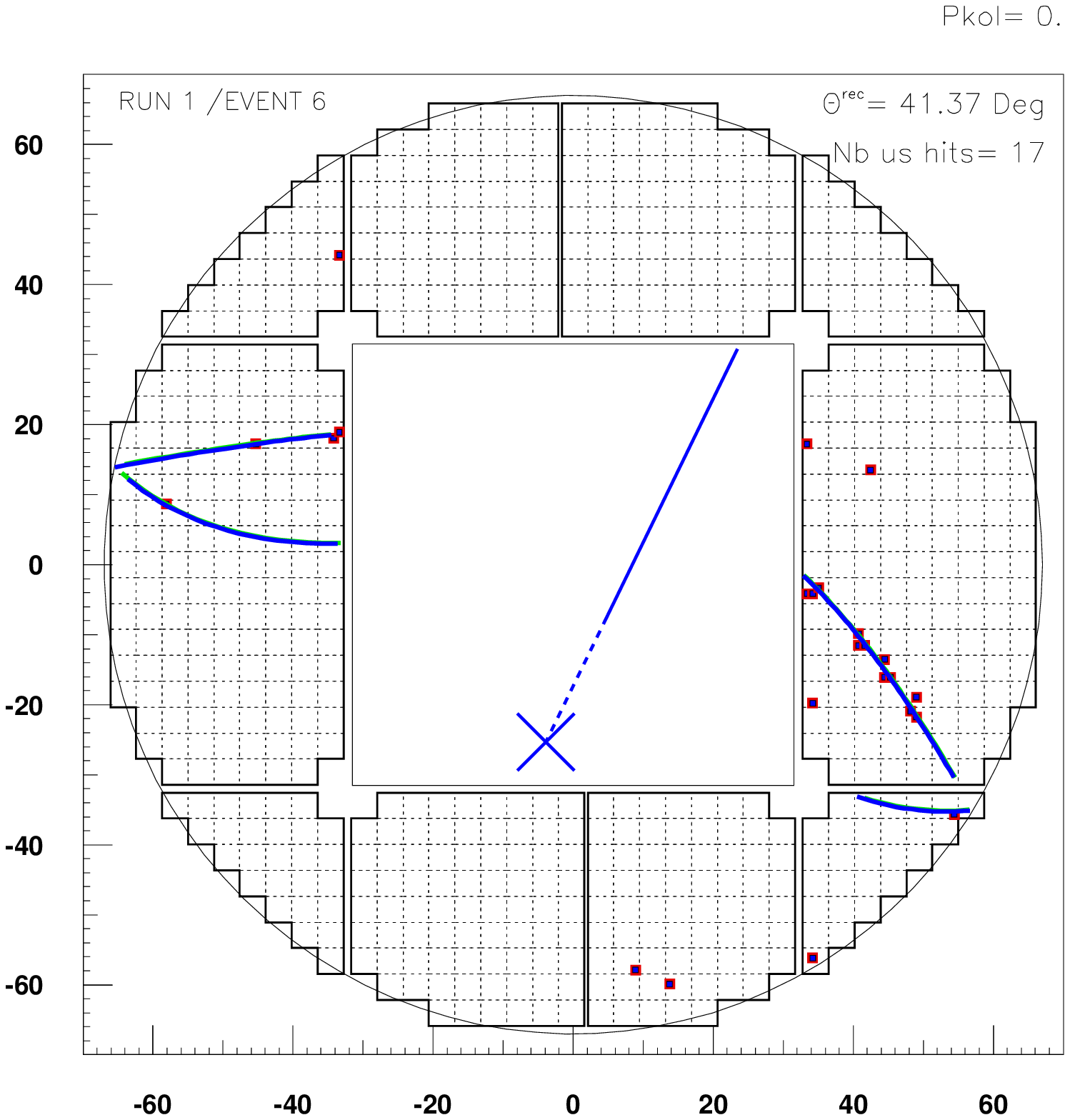} 
\end{tabular}
\caption{\emph{On the left:} View of the RICH detector. \emph{On the
  right:} Beryllium event display generated in a NaF radiator. The reconstructed photon pattern (full line)
  includes both reflected and non-reflected branches. The outer circular
  line corresponds to the lower 
boundary of the conical mirror. The square is the limit of the non-active
region.\label{fig:rich}}
\end{center} 
\end{figure}

\vspace{-1.8cm}
\emph{\subsection{The RICH detector}}
The RICH is a proximity focusing device with a dual radiator configuration on
the top (low refractive index aerogel 1.050, 3\,cm thick and a central square
of sodium fluoride (NaF), 0.5\,cm thick); a lateral conical mirror of 
high reflectivity increasing the 
reconstruction efficiency and a detection 
matrix with 680 photomultipliers and light guides. 
The active pixel size of the PMTs is planned to be of 8.5\,mm 
with a spectral response ranging from 300 to 650\,nm with a 
maximum at $\lambda\sim$\,420\,nm.
There will be a large non-active area at the centre of the detection 
area  due to the insertion of the ECAL. 
For a more detailed description of the RICH detector see 
Ref.~\refcite{bib:buenerd}.
The RICH detector of AMS was designed to measure the velocity of charged 
particles with a resolution $\Delta \beta/ \beta$ of 0.1\%, to extend the 
electric charge separation capability up to Z$\sim$26, to provide more information on the albedo rejection and to contribute in $e/p$ separation.
Its acceptance is of $\sim$0.35\,$m^2.sr$. 
Figure \ref{fig:rich} shows a view of the RICH and a beryllium event display
with a detailed view of the PMT matrix.

\vspace{0.3cm} 

\section{Velocity reconstruction}
A charged particle crossing a dielectric material of refractive index $n$
with a velocity $\beta$, greater than the speed of light in that 
medium, emits photons.
The aperture angle of the emitted photons with respect to the 
radiating particle is known as the \CK\ angle, $\theta_c$,
and it is given by (see Ref.~\refcite{bib:rich}).
\vspace{-0.1cm}
\begin{equation}
\cos\theta_c = \frac{1}{\beta~n}
\label{eq1}
\end{equation}
It follows that the velocity of the particle, $\beta$, is straightforward
derived from the \CK\ angle reconstruction, which is based on a fit to the
pattern of the detected photons.
Complex photon patterns can occur at the detector plane 
due to mirror reflected photons, as can be seen on right display of Fig.
\ref{fig:rich}. The event displayed is generated by a simulated beryllium nucleus in a
NaF radiator. 

\begin{figure}[htb]
\vspace{-0.1cm}
\begin{center}
\hspace{-.3cm}
\begin{tabular}{cc}
\hspace{-.5cm} 
\scalebox{0.40}{%
\begin{overpic}{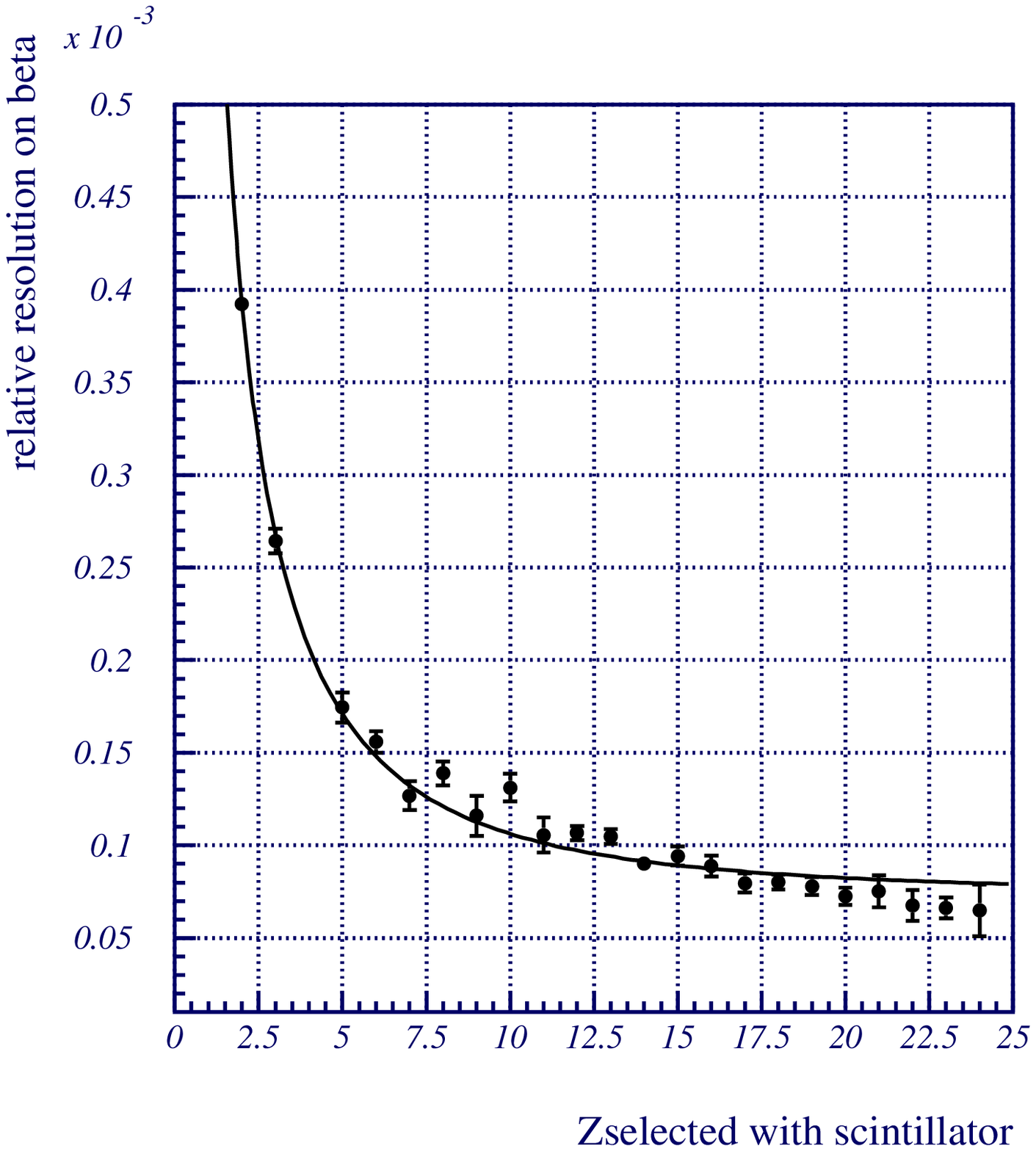} 
\put(32,75){\huge{$\sqrt {({\frac{A}{Z}})^{2}+B^{2}}$}}
\put(32,65){\huge{$A=7.77\times10^{-4} \pm 5\times10^{-6}$}}
\put(32,58){\huge{$B=7.3\times10^{-5} \pm 2\times10^{-6}$}}
\put(32,51){\huge{$\chi^{2}=2.76\qquad npfit=22$}}
\end{overpic}
}
&
\hspace{-1.3cm}
\scalebox{0.33}{%
\includegraphics[bb=-5 -5 485 507]{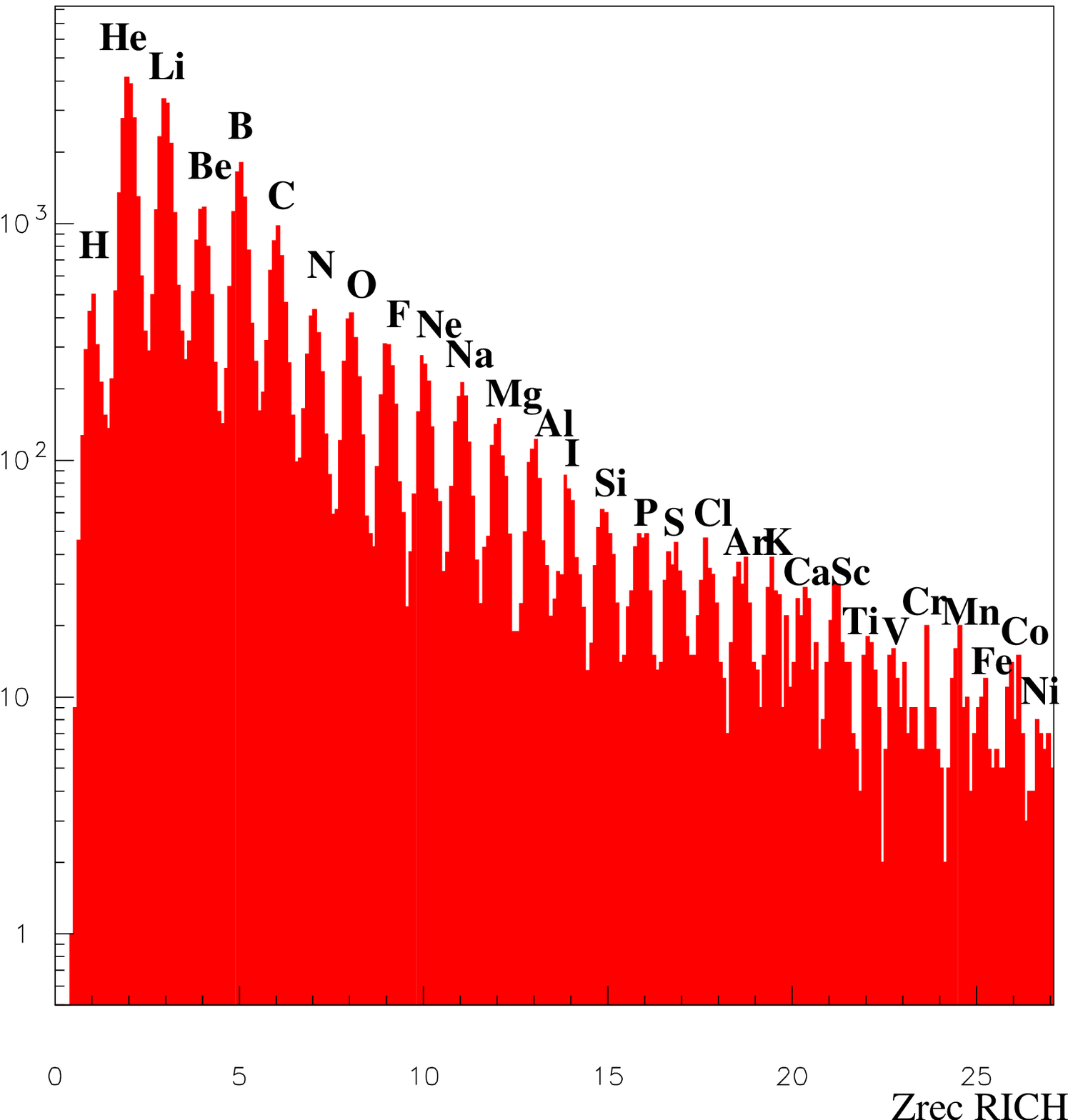} 
}                                                                          
\end{tabular} 
\vspace{-0.2cm}
\caption{At left evolution of the relative resolution on $\beta$ with the
  charge and at right the reconstructed charge peaks. Both are
  reconstructions with data from a test beam at CERN in October 2003, using an Indium beam of
158\,GeV/nuc. \label{fig:thc}}
\end{center}
\end{figure}

The \CK\ angle reconstruction procedure 
relies on the information of the particle direction 
provided by the Tracker.
The tagging of the hits signaling the passage 
of the particle through the solid light guides in the detection plane 
provides an additional track ele\-ment, however, those hits are excluded from the reconstruction.
The best value of $\theta_c$ will result from the maximization of a Likelihood function, 
built as the product of the probabilities, $p_i$, that the detected hits belong to a given (hypothetical)
\CK\ photon pattern ring,
\vspace{-0.2cm}
\begin{equation}
L(\theta_c) = \prod_{i=1}^{nhits} p_i^{n_i}  \left[ r_i(\theta_c) \right].
\label{eq:likelihood}
\vspace{-0.3cm}
\end{equation}
Here $r_i$ is the closest distance of the hit to the \CK\ pattern and $n_i$
the signal strength. For
a more complete description of the method see Ref.~\refcite{bib:NIM}.
The resolution achieved for singly charged particles crossing the aerogel
radiator with $\beta\sim$1 is $\sim$4\,mrad and for those crossing the NaF
radiator the resolution
is $\sim$8\,mrad. 
The evolution of the relative
resolution of $\beta$ with the charge can be observed on the left plot of
Fig. \ref{fig:thc}. It was extracted from reconstructed events generated in
a test beam at CERN in October 2003 with fragments of an Indium beam of
158\,GeV/nuc, in a prototype of the RICH detector.
\vspace{-0.8cm}
\vspace{-1.1cm}
\section{Charge reconstruction}
The \CK\ photons produced in the radiator are uniformly emitted along the particle path 
inside the dielectric medium, $L$, and their number per unit of energy 
depends on the particle's charge, $Z$, and velocity, $\beta$, and on 
the refractive index, $n$, according to the expression:
\begin{equation}
\frac{dN_{\gamma}}{dE} \propto Z^2 L \left( 1-\frac{1}{\beta^2 n^2} \right)=
Z^2 L \sin^2\theta_c
\label{eq:dnde}
\end{equation}
So to reconstruct the charge the following procedure is required:
\begin{itemize}
\vspace{-0.2cm}
\item \CK\ angle reconstruction. 
\vspace{-0.06cm}
\item Estimation of the particle path, $L$, which relies on the information of the particle 
direction provided by the Tracker.
\vspace{-0.06cm}
\item Counting the number of photoelectrons.
The number of photoelectrons related to the \CK\ ring has to be counted within a 
fiducial area, in order to exclude the uncorrelated background. Therefore, 
photons which are scattered in the radiator are excluded.  
A distance of 13\,mm to the ring was defined as the limit for photoelectron
counting, corresponding to a ring width of $\sim$5 pixels.
\vspace{-0.4cm}
\item Evaluation of the photon detection efficiency.
The number of radiated photons ($N_{\gamma}$) which will be detected ($n_{p.e}$) 
is reduced due to 
the interactions with the radiator ($\varepsilon_{rad}$), the photon ring acceptance 
($\varepsilon_{geo}$), light guide efficiency ($\varepsilon_{lg}$) and photomultiplier efficiency 
($\varepsilon_{pmt}$).
\vspace{-0.4cm}
\begin{equation}
n_{p.e.} \sim N_{\gamma}~\varepsilon_{rad}~\varepsilon_{geo}~\varepsilon_{lg}~\varepsilon_{pmt}  
\end{equation}
\end{itemize}
The charge is then calculated according to expression \ref{eq:dnde}, 
where the normalization constant can be evaluated from a calibrated beam of
charged particles.
Reconstructed charge
peaks are visible in the right plot of Fig.~\ref{fig:thc}. Data were
obtained with an aerogel radiator 1.05, 2.5\,cm thick from the mentioned test beam at CERN in
October 2003.
The charge resolution obtained for helium is $\sim$0.2 and it is possible to
separate charges up to Z=27.
For a more complete description of the charge reconstruction method see Ref.~\refcite{bib:NIM}.
\vspace{-0.2cm}
\section{Conclusions}
AMS is a spectrometer designed for antimatter and dark matter searches and for measuring
relative abundances of nuclei and isotopes.
The instrument will be equipped with a proximity focusing RICH detector based
on a mixed radiator of aerogel and sodium fluoride, enabling velocity measurements with a resolution of about 0.1\% and extending the charge measurements up to the Iron element.
Velocity reconstruction is made with a Likelihood method. Charge reconstruction is made in an event-by-event basis. 
Both algorithms were successfully applied to simulated data samples with flight configuration.
Evaluation of the algorithms on real data taken with the RICH prototype was
performed at the LPSC, Grenoble in 2001 and in the test
beam at CERN, in October 2002 and 2003.

\end{document}